%% file: main.tex
\documentclass[letterpaper]{article} %
\usepackage{aaai2027}  %
\usepackage[hyphens]{url}  %
\usepackage{graphicx} %
\usepackage{natbib}  %
\usepackage{caption} %
\usepackage{algorithm}
\usepackage{algorithmic}
\usepackage{enumitem}

\usepackage{newfloat}
\usepackage{listings}
\DeclareCaptionStyle{ruled}{labelfont=normalfont,labelsep=colon,strut=off} %
\floatstyle{ruled}
\newfloat{listing}{tb}{lst}{}
\floatname{listing}{Listing}

\usepackage{booktabs}
\usepackage{multirow}
\usepackage{url}
\usepackage{amsmath,amssymb} %
\usepackage{xspace}
\newcommand{\stopmethod}{\textsc{FailFast}\xspace}
\newcommand{\restartmethod}{\textsc{RestartSmart}\xspace}

\title{Fail-Fast, Restart-Smart:\\ Early Failure Prediction and Restart for SWE Agentic Tasks}

\nocopyright
\author{
    Chenyu Wang\textsuperscript{\rm 1},
    Yunbo Lyu\textsuperscript{\rm 1},
    Junda He\textsuperscript{\rm 1},
    Zhou Yang\textsuperscript{\rm 2},
    Chenxing Zhong\textsuperscript{\rm 3},
    Yaniv Harel\textsuperscript{\rm 4},
    David Lo\textsuperscript{\rm 1}
}
\affiliations{
    \textsuperscript{\rm 1}Singapore Management University\\
    \textsuperscript{\rm 2}University of Alberta\\
    \textsuperscript{\rm 3}Nanjing University of Science and Technology\\
    \textsuperscript{\rm 4}Tel Aviv University\\
    \{chenyuwang, yunbolyu, jundahe, davidlo\}@smu.edu.sg, zhou.yang@ualberta.ca,\\ chenxingzhong@njust.edu.cn, yaniv10@tauex.tau.ac.il
}

\begin{document}

\maketitle

\begin{abstract}
\input{sections/abstract}
\end{abstract}

\input{sections/introduction}

\input{sections/related_work}

\input{sections/method}

\input{sections/experiments}
\input{sections/conclusion}

\bibliography{references}

\clearpage
\appendix
\input{sections/appendix}

\end{document}

%% file: sections/abstract.tex
Software engineering (SWE) agents resolve repository-level issues through long trajectories that grow increasingly expensive as context accumulates.
Failed runs tend to be longer and exhibit redundant exploration or looping, suggesting that some failures may be detectable before completion.
Early termination, however, risks interrupting trajectories that would otherwise succeed; conversely, an unsuccessful trajectory may still contain useful repository edits.
We present \stopmethod--\restartmethod, a two-stage controller for a single active trajectory.\stopmethod is a lightweight $0.6$B monitor trained with terminal and dense fail-to-pass supervision to predict failure from observable prefixes without policy logits or hidden states.
Upon an alarm, \restartmethod launches a fresh same-policy rollout without prior prompt history and offers the interrupted repository diff as an optional overlay that the agent may inspect, apply, or discard.
On SWE-bench Verified, a monitor trained solely on Qwen3.6-27B trajectories transfers to three other policies, including a closed-API model, and saves $14.6\%$--$20.4\%$ of execution tokens at a target $5\%$ false-positive rate; on Qwen3.6-27B, its $20.4\%$ saving exceeds the $12.5\%$ achieved by our per-step AgentStop adaptation.
At a target $25\%$ false-positive rate, \restartmethod raises Qwen3.6-27B resolution from $66.6\%$ to $71.8\%$, whereas cold restart reaches only $66.8\%$.
Together, these results support early termination with sequential same-policy recovery.

%% file: sections/introduction.tex
\section{Introduction}
\label{sec:intro}

LLM-based coding systems have evolved from single-pass code generation into software engineering (SWE) agents that resolve repository-level issues by inspecting code, executing commands, editing files, and validating candidate patches~\citep{jimenez2024swe,yang2024sweagent,lyu2026practitioners}.
SWE agents operate through long interaction trajectories, and each step may reprocess an increasingly large history, producing the \emph{token snowball} effect~\citep{fan2025swe}.
Failed trajectories are especially costly: they tend to run longer and exhibit redundant exploration, repetition, or looping~\citep{majgaonkar2025understanding,bouzenia2025understanding}.
Although trajectory length alone is often a late signal, these recurring behaviors motivate the possibility of recognizing likely failure from an earlier prefix.

Early termination is nevertheless an intervention, not merely a prediction problem.
A false alarm can abort a trajectory that would otherwise succeed, so even an accurate failure predictor may reduce end-to-end performance~\citep{vasudev2026intervention}.
Conversely, terminal failure does not imply that all intermediate work is useless.
AgentLens classifies $54.9\%$ of its analyzed failing trajectories as \emph{Partial-fail}: trajectories that remain structurally close to known-good processes~\citep{sahoo2026agentlens}.
This suggests that some failed attempts may contain recoverable partial progress.
Together, these observations motivate our central question: \emph{Can we stop a likely-to-fail trajectory early and turn the saved compute and partial progress into a more successful retry?}

Prior work only partially addresses this question, largely treating early termination and recovery in isolation.
Post-hoc analyses reveal failure patterns but are not designed for online control~\citep{sahoo2026agentlens,majgaonkar2025understanding}, while online methods estimate when continued execution may be unproductive yet generally do not study recovery from the interrupted run~\citep{pham2026agentstop,baidya2026evidence,ruan2026doomed,guo2026eet}.
Recovery methods instead correct the ongoing trajectory~\citep{gandhi2025agents} or rely on multiple inference paths and stronger-model escalation~\citep{kim2026atropos}.
These approaches do not address a sequential same-policy retry: restarting from scratch discards potentially useful edits, whereas continuing the original trajectory may preserve the reasoning that led it astray.
The open question is therefore not only when to stop, but what, if anything, should cross the restart boundary so that recovery can outweigh false-positive harm and retry cost.

To address this gap, we propose \emph{\stopmethod--\restartmethod}, a controller for a single active trajectory that combines per-step failure prediction with edit-preserving same-policy retry.
\stopmethod is a lightweight $0.6$B monitor that reads the issue statement and a window of recent thoughts, actions, and observations.
We augment terminal-outcome supervision with dense fail-to-pass (F2P) progress targets obtained by replaying agent commands and evaluating intermediate patches.
Because FailFast relies only on observable text rather than policy logits or hidden states, it can monitor both open-weight and closed-API policies after threshold calibration \emph{on validation data} under a target false-positive-rate budget.
When an alarm fires, \restartmethod launches a fresh retry without the previous prompt history and exposes a stable repository diff as an optional, unverified, \texttt{git-apply}-backed overlay.
The overlay starts disabled and may be inspected, applied, or discarded, separating potentially useful edits from the reasoning that produced them.

Experiments on SWE-bench Verified reveal complementary cost-saving and recovery regimes.
Trained solely on Qwen3.6-27B trajectories, \stopmethod transfers without weight updates to three other policies, including Gemini 3 Flash.
At a target $5\%$ false-positive rate (FPR)---i.e., stopping approximately $5\%$ of otherwise successful runs---it saves $14.6\%$--$20.4\%$ of total execution tokens across the four policies; on Qwen3.6-27B, its $20.4\%$ saving exceeds AgentStop's $12.5\%$ and a duration-only control's $11.4\%$.
Including monitor overhead, this corresponds to estimated inference-energy and operational-carbon savings of $14.5\%$--$20.3\%$ across the three open-weight policies.
Because the $5\%$ operating point stops too few trajectories for effective recovery, we evaluate \restartmethod at target FPRs of $10\%$ and $25\%$.
At $10\%$, it improves resolution by $3.0$--$4.0$ percentage points across three open-weight policies, with $20.5\%$--$36.9\%$ net token overhead.
At $25\%$, it raises Qwen3.6-27B's resolution from $66.6\%$ to $71.8\%$ ($+5.2$ percentage points), whereas cold restart under the same alarms reaches only $66.8\%$ ($+0.2$ points).
The overhead is thus a fraction of a single additional rollout, and well below the sampling budgets that test-time scaling commits up front~\citep{jain2025r2egymproceduralenvironmentshybrid,pmlr-v267-pan25g}.

%% file: sections/related_work.tex
\section{Related Work}
\label{sec:related_work}

Improving SWE-agent outcomes at test time means deciding how much compute to spend and where.
Prior work answers by sampling more, by analyzing failure offline, or by stopping a run online, but rarely by connecting the stopping decision to what happens next.

\textbf{Sampling more.}
The dominant way to improve test-time performance is to draw multiple full trajectories and select among them, whether by verifier, rubric, or search~\citep{pmlr-v267-pan25g,jain2025r2egymproceduralenvironmentshybrid,raghavendra2026agenticrubricscontextualverifiers,shum2025swerm,gao2025trae,antoniades2025swe,aggarwal2025dars,lin2025seagent,ding2026swereplay,han2026swetrace}, or by supervising the candidate pool online~\citep{lin2025qlass,xi2025agentprm,chen2025birm,park2025calibration,aggarwal2023adaptiveconsistency,li2024esc,sun2024speculative,wang2026criticrubrics}.
These methods commit multiples of one trajectory regardless of instance difficulty, and all require a pool of candidates that a single deployed run does not have.

\textbf{Failure signals exist but are used offline.}
Trajectory analyses show that failure is often detectable before termination: failed runs are longer and more variable, exhibit recurring anti-patterns, and accumulate execution errors~\citep{majgaonkar2025understanding,bouzenia2025understanding,chen2026beyondfinalcode,mehtiyev2026beyondresolution,ceka2025traceability,cemri2025mast,zhang2025whowhen,barke2026agentrx,fan2026agentprocessbench}.
They operate offline, however, or score trajectories against references derived from multiple passing runs of the same issue, which are unavailable for an unseen issue at deployment~\citep{sahoo2026agentlens}.
We convert these prefix signals into a calibrated online predictor instead.

\textbf{Stopping and recovery, studied apart.}
Single-run early stoppers terminate likely-to-fail trajectories online, but evaluate detection and savings alone, and rely on signals a deployed run may not expose: log-probability supervisors need white-box access to the policy~\citep{pham2026agentstop}, as do hidden-state probes~\citep{ruan2026doomed}; weak supervision targets sparse, late-arriving evidence~\citep{baidya2026evidence}; experience-conditioned confidence gates decisions around a completed patch rather than mid-trajectory~\citep{guo2026eet,zhang2026agentforesight}.
None reinvests what it reclaims.
Recovery methods instead modify the run in place, through live feedback~\citep{gandhi2025agents,nanda2026wink} or step rollback~\citep{liu2026process,li2025garollback}, but correction is not cheap: feedback requires a monitor stronger than the policy~\citep{gandhi2025agents}, rollback a policy-scale assistant~\citep{li2025garollback}, and our reproduction shows that the prompt intervention improving a weak agent reduces the resolution rate of a strong one (Section~\ref{sec:experiments:results}).
\citet{kim2026atropos} are closest, pairing prediction with recovery, but escalate to a stronger model under merged parallel rollouts.
The remaining case is a sequential retry under the same policy: what, if anything, should carry from an aborted run into its replacement.

\textbf{Why prediction alone is not enough.}
Agents' budget forecasts are over-optimistic~\citep{lin2026bagen}, so stopping requires an external monitor; and accurate prediction does not imply beneficial intervention, since any intervention both recovers failing runs and disrupts runs that would have passed~\citep{vasudev2026intervention}.
Our design addresses both: the monitor operates under an explicit false-positive budget, and the restart offers rather than imposes the salvaged edits.

%% file: sections/method.tex
\section{Method}
\label{sec:method}

We study whether an agent's SWE-bench trajectory can be judged \emph{online} from its partial prefix, before it terminates.
Such a judgement enables a fail-fast-and-restart controller: abandon a run predicted to fail rather than pay for its full step budget.
We formulate early termination as per-step failure prediction and cast the deployment decision as a budgeted alarm: at a fixed false-positive rate (FPR, the fraction of \emph{successful} runs wrongly aborted), maximize recall (the fraction of \emph{failing} runs caught).
We formulate restart as information extraction: given a stopped trajectory, what partial progress can be extracted and exposed to a fresh rollout?
The controller has two components.
\emph{(i) Early termination (\stopmethod)}: a monitor that decides \emph{when} to abort a single run.
\emph{(ii) Restart policy (\restartmethod)}: a mechanism that decides \emph{what to carry over} from the aborted trajectory into the retry.

\subsection{Trajectory Early Termination: \stopmethod}
\label{sec:method:earlystop}

Because prior studies lack a dedicated method for effectively early-terminating a single trajectory that is likely to fail (without access to ground truth or parallel samples from the same instance), we design a custom monitor.
The monitor is a language model with lightweight probing heads, trained on trajectories produced by the evaluated policy model (Qwen3.6-27B-FP8, sampled using seeds 0 through 10 in fixed environments) on SWE-bench-verified~\cite{jimenez2024swe, chowdhury2024swebenchverified} under mini-swe-agent~\cite{yang2024sweagent}, a lightweight framework whose evaluated models closely match their officially reported SWE-bench, and has been used in many previous works~\cite{guo2026eet, lyu2026agentszz,ding2026swereplay}.

\subsubsection{State Representation and Auxiliary Targets}
\label{sec:method:state_targets}

Each trajectory is a sequence of steps $t=0,1,\dots,T$ that receives a terminal label $y_{\text{final}}\!\in\!\{0,1\}$ indicating ultimate success. 
A training sample is a trajectory prefix truncated at step $t$.
We serialize the prefix into two components containing strictly online information: (1)~\texttt{[ISSUE]}, the task specification, and (2)~\texttt{[WINDOW]}, the recent eight steps of thought, action, and observation, plus a persistent pin of the latest patch-producing step. 

Relying solely on a single binary label at the end of a long trajectory provides sparse, delayed supervision, which severely exacerbates the credit assignment problem when evaluating intermediate steps. To mitigate this and provide denser, fine-grained supervision, we attach two progress targets $g^{\text{f2p}}_t, g^{\text{p2p}}_t \in [0,1]$ to each prefix.
We compute these by replaying the prefix's bash commands in a fresh instance environment to extract an intermediate patch, which is then scored by the official terminal evaluator for \emph{fail-to-pass} and \emph{pass-to-pass} pass rates.
These targets initialize at $(0,1)$---reflecting the unmodified repository where all \emph{pass-to-pass} tests succeed but \emph{fail-to-pass} tests fail---and update only upon valid edits. For non-editing steps or malformed patches, we carry forward the last valid score, explicitly resetting to $(0,1)$ if the agent reverts the working tree.
We also construct Bradley--Terry~\cite{BradleyTerry1952} preference pairs by pairing successful and failed prefixes from mixed instances, matched within each step-fraction decile bucket. 
This pairing controls for both instance difficulty and trajectory progress, isolating an instance-invariant ranking signal.

\subsubsection{Monitor architecture and training}
\label{sec:method:arch}
The monitor backbone is a frozen language model adapted via LoRA~\cite{hu2021loralowrankadaptationlarge} on attention and MLP projections. 
Over the pooled last-token hidden state, we attach three linear heads kept in \texttt{fp32}: a value head $s$ for ultimate resolution, an F2P head $\hat{a}_f$, and a P2P head $\hat{a}_p$.
Only the LoRA adapter and heads are trained.

The training objective minimizes the binary cross-entropy (BCE) of the value head $s$ against the terminal label, alongside auxiliary progress and ranking terms:
\begin{equation}
\begin{aligned}
\mathcal{L} = {}& \mathrm{BCE}\!\big(s,\, y_{\text{final}}\big)
+ \lambda_f\,\mathrm{BCE}\!\big(\hat a_f,\, g^{\text{f2p}}\big) \\
&+ \lambda_p\,\mathrm{BCE}\!\big(\hat a_p,\, g^{\text{p2p}}\big)
+ \lambda_r\big[{-\log\sigma(s_{+}-s_{-})}\big] .
\end{aligned}
\end{equation}
This regresses both the global decision target and the dense progress features. We empirically set $\lambda_p=0$, as subsequent experiments demonstrate that the F2P signal yields consistent improvements while the P2P signal does not.

\subsubsection{Calibration and the deployment policy}
\label{sec:method:policy}
For deployment, the raw value logit is calibrated to probabilities using Platt scaling~\cite{platt1999probabilistic}.
To construct a per-step decision rule, we employ a \emph{meta score}, $p_{\text{fail}} = \sigma\!\big(w^\top[\,v,\ \hat g_{f},\ \phi\,]+b\big)$, a logistic regression fusing the calibrated value $v$, the F2P progress estimate $\hat g_{f}$, and the step fraction $\phi$.
This consistently outperforms a purely value-based threshold by effectively penalizing low-progress, late-stage states.
The trajectory-level alarm aggregates per-step meta scores. 
To prevent premature firing on the noisy opening steps, decisions are gated by a minimum step fraction $\phi \geq \phi_{\min}$, which carries the added benefit of completely bypassing monitor inference during early steps (experiments showed that $\phi_{\min}=20$ steps provides the best balance).
The monitor triggers an abort either when cumulative votes reach a threshold $M$ (\emph{count-M}) or after $K$ consecutive votes (\emph{sustained-K}).
We select the deployment operating point by a budget-constrained grid search, maximizing recall subject to the target false-positive rate.
The monitor is trained entirely out-of-fold; only calibration and threshold selection are fitted on evaluation-fold predictions.
A nested instance-level cross-fit yields nearly identical operating points and performance, suggesting the selected policy is not an artifact of test-label overfitting.

\subsection{Restart Policy: \restartmethod}
\label{sec:method:restart}
 
When the monitor aborts a run at step $t_f$, it halts the trajectory and triggers \restartmethod, an edit-preserving restart policy operating with a fresh budget.
While escalating aborted instances to a larger, more capable model is an intuitive option, this study specifically investigates whether the \emph{same} policy model can utilize artifacts distilled from its own early-stopped trajectory to recover from failure.
Outperforming an unguided re-sample (cold restart) under the same model proves that our distilled artifacts extract genuinely constructive guidance from doomed runs.
A natural first instinct is to summarize the aborted trajectory into textual prompt context; however, we notice that verbose text summaries induce severe LLM \emph{anchoring effects}~\cite{huang2026understanding}, prematurely locking the restarting agent into the prior attempt's faulty reasoning paths.
To circumvent textual anchoring, we instead preserve the agent's physical code edits as a clean, environment-level overlay: a fresh rollout inherits no prior prompt history, but is warm-started with access to the aborted repository edits, which it is free to inspect, adopt, or discard.

\paragraph{Extracting the past edits.}
We replay the aborted trajectory's bash commands, capturing \texttt{git diff HEAD} after each command and recording a snapshot whenever the diff changes, restricted to files existing at the task baseline (files newly created by the agent are excluded). 
Crucially, an abort often interrupts the agent just as it is about to apply an edit or while it is mid-repair.
To capture a logically complete patch, we allow the replay to finish the agent's intended sequence of modifications before snapshotting.
We therefore do not cut exactly at $t_f$: we locate the first edit at or after $t_f$ (falling back to the most recent prior edit if none exist), and extend the cut until $g$ consecutive steps pass without any further edits.
The overlay $\Delta$ is the cumulative diff at this cut, which is highly likely to be a settled edit rather than a mid-edit fragment.

\paragraph{Carrying edits as an overlay.}
When restarting these early-stopped trajectories, we expose $\Delta$ as an \emph{overlay}: \emph{offered, not forced}.
Rather than splicing it into the prompt, we mount it as a removable, git-apply--backed tool the agent may inspect ($\texttt{diff}$), apply ($\texttt{on}$), or revert ($\texttt{off}$); the overlay starts \emph{off}, and the prompt frames it as an \emph{unverified} lead from a stopped attempt.
This lets a restart skip rediscovery costs when it trusts the prior work, or rebuild from scratch when it does not, outperforming cold restarts.

%% file: sections/experiments.tex
\section{Experiments}
\label{sec:experiments}

\subsection{Setup}
\label{sec:experiments:setup}

\paragraph{Dataset and Task.}
Following similar trajectory-collection setups for SWE verifiers~\cite{pmlr-v267-pan25g, shum2025swerm}, we collect $11$ independent runs (seeds $0$–$10$) over $500$ SWE-bench Verified instances, partitioned into $350$ training, $50$ validation, and $100$ test instances.
For the standalone evaluation of \stopmethod, we train the monitor only on \texttt{Qwen3.6-27B} trajectories and evaluate it across different policy models, testing whether it captures general failure patterns rather than model-specific behavior.
For the \restartmethod evaluation, we instead use a policy-specific monitor trained on trajectories from the corresponding model, allowing us to assess the potential of restart under the most suitable monitor.
Each trajectory contains at most 100 steps, following SWE-bench Leaderboard~\cite{openlm_swebench_2026}.

\paragraph{Models and Baselines.}
To minimize online inference overhead, we deploy a \texttt{Qwen3-0.6B} monitor backbone (occupying only $\sim$2GB VRAM including both model and context).
This single monitor model is trained exclusively on the \texttt{Qwen3.6-27B} trajectory pool and deployed across four policy models, spanning three state-of-the-art open-weight LLMs highly capable at their respective scale tiers---\texttt{Qwen3.5-9B}, \texttt{Qwen3.6-27B}, and \texttt{Gemma4-31B}---and one popular proprietary LLM, \texttt{Gemini3-Flash} (natively resolving 48\%, 66\%, 62\%, and 67\% of SWE-bench Verified tasks respectively).
Scaling the monitor backbone beyond $0.6$B yields no statistically significant performance gain (Section~\ref{para:arch}), keeping monitoring overhead minimal relative to the policy model.

AgentStop~\citep{pham2026agentstop} predicts failure by feeding policy execution signals (token log-probabilities, token counts, and step repetitions) into a gradient-boosted tree.
In its original form, the decision is made once at a predetermined checkpoint $k$ ($k \le 10$): a step-$k$ classifier inspects the first $k$ steps, leaving surviving trajectories unchecked thereafter.
We reproduce this design from the authors' released code and train it on the same trajectories and labels as our monitor.
To adapt AgentStop to our per-step setting, we extend the checkpoint sweep across the entire run, which substantially improves recall across all operating points while reclaiming a higher fraction of tokens in most cases.

To isolate performance beyond trajectory length alone, we also include a supervisor-free \emph{Duration} control.
Because failing trajectories tend to run longer than successful ones~\citep{majgaonkar2025understanding}, this control scores runs solely using step count, calibrated and thresholded across FPR budgets identically to learned monitors.
It strictly upper-bounds signal-free stopping and substantially dominates the random-exit baseline evaluated in \citet{pham2026agentstop}.

Studies about restarting single-trajectory aborted are scarce in software engineering agents.
We thus broaden our scope to non-voting single-trajectory methods (excluding multi-sample setups like Best-of-$N$), selecting \textbf{SWE-PRM}~\citep{gandhi2025agents} as our primary baseline.
SWE-PRM is an inference-time Process Reward Model that detects and course-corrects errors via taxonomy-guided prompt feedback; we use its top-performing setup with \texttt{Claude-Sonnet-4}.
Additionally, we evaluate a \emph{Cold restart} control, which re-initiates tasks aborted by \stopmethod anew without prior context.
While we initially considered \citet{liu2026process}, its evaluation is restricted to errors persisting across repeated runs.
In our preliminary tests, it slightly reduced the overall resolve rate, likely by disrupting successful trajectories, so we exclude it from the main comparison.

\paragraph{Training and Hyperparameters.}
The monitor backbone is adapted via LoRA~\cite{hu2021loralowrankadaptationlarge} with rank $r\!=\!16$ and scaling factor $\alpha\!=\!32$.
Although the training objective admits a P2P progress term for symmetry with F2P, we found it to yield no significant gain and set $\lambda_p=0$ in the deployed configuration.
The final deployed configuration, determined empirically, uses $\lambda_f=0.5$, $\lambda_r=0.25$, $\lambda_p=0$.

\subsection{Main Results}
\label{sec:experiments:results}

Metrics are averaged over 5 independent runs (seeds 0--4).

\subsubsection{Early-Stop Monitor (\stopmethod) Performance.}
\label{sec:experiments:results:earlystop}

Table~\ref{tab:earlystop} summarizes monitor performance.
Practically, the \textbf{FPR budget} controls the system's collateral damage tolerance, i.e., the maximum fraction of would-pass trajectories allowed to be accidentally aborted.
A tight budget ($5\%$ FPR) minimizes accuracy loss for pure cost saving, while a looser budget (e.g., $25\%$ FPR) tolerates more false alarms to maximize failure detection for subsequent restarts (re-running aborted tasks utilizing prior trajectory context).
Under each budget, \textbf{Recall} reflects the monitor's detection power: the proportion of would-fail runs caught and aborted before full execution.
\textbf{Precision} measures alarm reliability (true failures among aborted runs), \textbf{Fired} is the percentage of test trajectories stopped, and \textbf{Saved} is the fraction of total agent compute (in tokens) reclaimed by early abortion, relative to running all trajectories (aborted and completed) to completion.

\input{tables/tab_earlystop.tex}

\textbf{\stopmethod substantially reduces compute budget and transfers across model families.}
Across all FPR budgets and policy models (Table~\ref{tab:earlystop}), \stopmethod reclaims the highest percentage of total agent compute.
For instance, on \texttt{Qwen3.6-27B}, when overall performance degrades by at most $5\%$ (FPR budget), \stopmethod salvages over a fifth (${20.4\%}$) of all token compute, substantially outperforming AgentStop ($12.5\%$) and the Duration control ($11.4\%$).
At a looser $25\%$ FPR, \stopmethod intercepts $68.3\%$ of all failing runs while reclaiming ${49.0\%}$ of total tokens (versus around $33\%$--$37\%$ for the baselines), providing a substantial compute buffer to subsidize subsequent restarts for these intercepted cases.
Across all three policy models at the $5\%$ FPR budget, this single 27B-trained monitor consistently salvages $15\%$--$20\%$ of total token costs regardless of the target model.

\textbf{\stopmethod introduces negligible estimated energy and carbon overhead.}
Because \stopmethod deploys an LLM-based monitor (occupying $<2$GB VRAM), it inherently incurs inference overhead, prompting us to examine whether raw token savings translate into net energy and carbon reductions.
Per-token compute scales with parameter count \(N\) at approximately \(2N\) FLOPs/token~\citep{kaplan2020scaling}, making our 0.6B monitor \(15\times\)--\(52\times\) cheaper per token than the 9--31B policy models.
The monitor adds only a $0.1\%$--$0.6\%$ estimated compute overhead.
Under the standard approximation that inference energy scales with model FLOPs, the \textbf{Saved} fraction in Table~\ref{tab:earlystop} provides a hardware-agnostic estimate of the policy-inference energy reclaimed by early termination.
After subtracting the monitor overhead, the resulting values approximate the net energy reduction.
They also approximate carbon reductions when the monitor and policy inference are executed under comparable hardware, data-center efficiency, and grid carbon intensity.

\textbf{\stopmethod accurately detects failures and transfers across model families.}
\stopmethod outperforms all baselines across policies and operating points, with AgentStop holding a marginal edge (around $0.5\%$ recall) in only two configurations.
On the native \texttt{Qwen3.6-27B} agent it achieves $30.5\%$ recall at $5\%$ FPR and $68.3\%$ at $25\%$, exceeding AgentStop by up to $14$ percentage points; transferring across families to \texttt{Gemma4-31B} yields $27.2\%$ and $57.4\%$, against $23.8\%$ and $51.5\%$ for AgentStop.
On \texttt{Qwen3.5-9B} all methods reach comparable recall at $5\%$ FPR ($34.7\%$, $35.1\%$, $34.7\%$), likely because a few extremely long failures on weaker policies are already caught by step-count rules.
Recall alone, however, is not enough: Duration and AgentStop fire later and reclaim only ${\sim}10\%$ of tokens, whereas \stopmethod reclaims $15.7\%$ at the same recall.
At $25\%$ FPR it flags $72.2\%$ of failing runs and saves $50.1\%$ of tokens, leaving headroom for restarts.

\paragraph{Transfer to Closed-Source APIs.}
\input{tables/tab_gemini.tex}
We further evaluate whether \stopmethod generalizes to closed-source APIs using \texttt{Gemini3-Flash}. Note that AgentStop is structurally inapplicable here, as most proprietary APIs (especially those with thinking enabled) do not expose internal token probabilities.
As shown in Table~\ref{tab:gemini}, although detection recall experiences a slight decrease compared to open-weight models ($22.7\%$ at $5\%$ FPR and $44.8\%$ at $25\%$ FPR), the token savings remain firmly in the top tier: \stopmethod salvages ${16.0\%}$ of total token compute under a strict $5\%$ FPR budget (${37.1\%}$ at $25\%$ FPR), consistently outperforming the Duration control ($12.0\%$ and $30.2\%$, respectively).

\subsubsection{Restart Policy (\restartmethod) Performance.}
\label{sec:experiments:results:restart}
Table~\ref{tab:restart} evaluates \restartmethod on trajectories aborted by \stopmethod, where each model uses a monitor trained on its own trajectories.
Because the $5\%$ FPR budget intercepts too few runs to serve as an effective restart threshold, we focus on the $10\%$ and $25\%$ FPR operating points.
We compare \restartmethod against cold restart and the prompt-intervention baseline SWE-PRM~\citep{gandhi2025agents}.

\input{tables/tab_restart.tex}

\textbf{\restartmethod outperforms cold restart by preserving trajectory context.}
By warm-starting fresh rollouts with the prior edit overlay, \restartmethod consistently translates early-stop alarms into overall resolve rate gains—boosting \texttt{Qwen3.5-9B} up to $52.2\%$ ($+4.0\%$), \texttt{Qwen3.6-27B} up to $71.8\%$ ($+5.2\%$), and \texttt{Gemma4-31B} to $65.2\%$ ($+3.0\%$).
In contrast, \emph{Cold restart} achieves only marginal gains at the $10\%$ FPR budget ($+2.6\%$ for \texttt{Qwen3.5-9B}, $+0.6\%$ for \texttt{Qwen3.6-27B} and $+1.8\%$ for \texttt{Gemma4-31B}) that shrink further at $25\%$ FPR ($+1.6\%$, $+0.2\%$ and $+1.4\%$).
Comparing the two strategies demonstrates the necessity of trajectory context: across operating points, \restartmethod recovers significantly more genuine failures (\emph{TP rec.} up to $28.9\%$ vs.\ $20.2\%$ for cold restart on \texttt{Qwen3.6-27B}, and $21.7\%$ vs.\ $16.8\%$ on \texttt{Qwen3.5-9B}).
Crucially, the larger and more capable \texttt{Qwen3.6-27B} agent exhibits a markedly superior ability to leverage the overlay—critically evaluating prior modifications to preserve viable edits and restore false-alarm runs back to completion, keeping \emph{FP lost} as low as $8.8\%$ at $25\%$ FPR (vs.\ $27.5\%$ for cold restart, and $15.2\%$ vs.\ $36.4\%$ at $10\%$ FPR).
In contrast, without prior context, cold restart severely damages false alarms—destroying up to $36.4\%$ on \texttt{Qwen3.6-27B}—causing the penalty of discarding successful attempts to quickly outweigh any recovery gains.

In terms of token expenditure, cold restart incurs a lower total token cost than \restartmethod. Because cold restart requires no context from early-stopped trajectories, it can terminate execution immediately at the step flagged by \stopmethod.
In contrast, \restartmethod must wait for subsequent edits following the early-stop signal to construct the edit overlay, resulting in a higher token footprint (e.g., $+43.8\%$ vs.\ $+18.2\%$ for \texttt{Qwen3.6-27B} at $25\%$ FPR). Nevertheless, this extra investment is well justified, yielding a net $+5.0\%$ gain in overall resolve rate over cold restart ($+5.2\%$ vs.\ $+0.2\%$).

\textbf{Comparison with Prompt Intervention. (SWE-PRM)}
For the weaker \texttt{Qwen3.5-9B} agent, this prompt-intervention strategy proves effective, boosting the overall resolve rate by $4.2\%$ (performing on par with our restart method's $+4.0\%$ peak). 
However, this success does not transfer to significantly stronger policy models.
On the highly capable \texttt{Qwen3.6-27B} agent, SWE-PRM backfires, reducing the resolve rate by $3.2\%$.
This aligns with our expectation that while weaker models can be improved by explicit prompt guidance provided by much stronger models, stronger models cannot.
For highly capable policy models, intrusive prompt interventions often disrupt their coherent reasoning and problem-solving strategies, leading to worse performance. Furthermore, SWE-PRM incurs additional inference overhead and lacks a mechanism to explicitly control the false positive rate (FPR) budget. 
Our approach avoids these pitfalls by aborting and restarting the trajectory, preserving the agent's autonomy while recovering from failures.

\textbf{Restarts offer a controllable compute-performance tradeoff.}
Executing a restart inherently spends extra steps, partially offset by the tokens reclaimed from early stopping. 
For the \texttt{Qwen3.6-27B} agent, maximizing the resolve rate ($+5.2\%$) at the $25\%$ FPR budget requires investing roughly $+43.8\%$ net compute.
However, the $10\%$ FPR operating point provides a more compute-efficient tradeoff for \texttt{Qwen3.6-27B}, trading fewer, more-confident alarms for a cheaper gain: achieving $+3.2\%$ resolve for $+30.3\%$ extra compute.
For the \texttt{Qwen3.5-9B} agent, the $10\%$ FPR budget acts as a sweet spot, achieving the peak $+4.0\%$ resolve rate for $+36.9\%$ net compute, whereas pushing to $25\%$ FPR yields no further gains ($+3.8\%$) while increasing the cost ($+39.2\%$).
Notably, under cold restart, \texttt{Qwen3.5-9B} incurs an identical net token cost ($+12.1\%$) at both $10\%$ and $25\%$ FPR thresholds; the earlier termination at $25\%$ FPR reclaims higher per-run token savings that precisely offset the overhead of interrupting a larger volume of instances.
This flexibility allows users to adjust the FPR threshold depending on their available compute budget and policy model.

\subsection{Ablation Study and Analyses}
\label{sec:experiments:ablations}
We conduct ablation studies for \stopmethod and \restartmethod using monitors trained on trajectory datasets from their respective monitored policy models (\texttt{Qwen3.5-9B} and \texttt{Qwen3.6-27B}).

\subsubsection{\stopmethod}
\label{sec:experiments:ablations:earlystop}

\paragraph{Influence of Monitor Backbone Capacity.}
\label{para:arch}

\input{tables/tab_arch.tex}

To evaluate the impact of backbone capacity, we compare our lightweight \texttt{Qwen3-0.6B} monitor against a larger \texttt{Qwen3-4B} model across FPR budgets (Table~\ref{tab:arch}).
Scaling the monitor backbone from 0.6B to 4B parameters yields no meaningful performance gain.
Across both policy models, \texttt{Qwen3-0.6B} achieves comparable or even superior recall compared to \texttt{Qwen3-4B} (e.g., $45.5\%$ vs.\ $41.3\%$ at $10\%$ FPR on \texttt{Qwen3.6-27B}, and $55.2\%$ vs.\ $51.7\%$ on \texttt{Qwen3.5-9B}).
This aligns with scaling bottlenecks in scalar reward modeling~\citep{chen2025rm}, showing that prefix-level failure signals saturate early—allowing our ultra-lightweight \texttt{Qwen3-0.6B} monitor to deliver peak efficacy at near-zero inference cost.

\paragraph{Influence of Input Features.}
\label{para:input_ablation}

\input{tables/tab_input_ablation.tex}

The monitor prompt can incorporate three candidate input components designed to capture failure signals at different granularities: (1)~\texttt{I} (\texttt{[ISSUE]}) conveys task intent and inherent problem difficulty; (2)~\texttt{W} (\texttt{[WINDOW]}) exposes recent action-observation history to detect immediate failure symptoms; and (3)~\texttt{S} (\texttt{[STATE]}) provides a global trajectory-quality summary from step one, adapting the process-centric abstraction of \citet{liu2026process}.
Table~\ref{tab:input_ablation} evaluates their contributions by comparing our \emph{Default} configuration (\texttt{I+W}) against omitting components (\texttt{$-$I}, \texttt{$-$W}) or appending global state (\texttt{+S}).
First, adding the global process summary (\texttt{+S}) yields no benefit and actually degrades performance in most cases (e.g., dropping from $58.1\%$ to $55.1\%$ at $15\%$ FPR on \texttt{Qwen3.6-27B}).
This indicates that coarse heuristic summaries act as distractors when an LM monitor can directly analyze raw execution text, justifying our choice to omit S by default.
Second, recent execution history (W) is the single most critical feature; omitting it (\texttt{$-$W}) causes the sharpest recall drop across policy models (e.g., falling by $13.2\%$ from $58.1\%$ to $44.9\%$ at $15\%$ FPR on \texttt{Qwen3.6-27B}).
This confirms that \stopmethod's early termination decisions depend tightly on dynamic trajectory behavior rather than static issue descriptions alone.
Omitting problem context (\texttt{$-$I}) also degrades performance, showing that task intent provides essential baseline context.

\paragraph{Additional Ablations for \stopmethod.}
The supplementary material reports three additional analyses to verify that the \stopmethod design and training settings adopted in this paper are appropriate.
First, we vary the observation-window size, i.e., the number of recent trajectory steps included in the monitor input, to determine how much execution history is needed.
Second, we compare training with only the value head against training with additional F2P and P2P heads to examine whether dense progress supervision improves the value head used for early-stop decisions.
Third, we vary the number of independently sampled policy trajectories per instance to study how trajectory diversity affects monitor performance and to validate the selected training-data scale.

\subsubsection{\restartmethod}
\label{sec:experiments:ablations:restart}

\paragraph{Influence of Edit-Completion Patience.}

\input{tables/tab_restart_patience.tex}

To avoid capturing incomplete or pending code edits, \restartmethod locates the first edit at or after the abort signal and delays termination until five steps pass without further edits (Table~\ref{tab:restart_patience}).
An immediate cut ignores this patience window, inheriting fragmented overlays that wrongly destroy viable runs (e.g., $18.8\%$ FP lost vs.\ $8.8\%$ at $25\%$ FPR).
By allowing the target edit to fully settle, our \emph{Wait} policy preserves complete, coherent overlays, boosting the final resolve rate to $71.8\%$ ($+5.2\%$).

%% file: tables/tab_earlystop.tex
\begin{table*}[!t]
\centering
\small
\setlength{\tabcolsep}{4.5pt}
\begin{tabular}{@{} l l cccc cccc cccc @{}}
\toprule
\textbf{Policy Model} & \textbf{FPR} & \multicolumn{4}{c}{\textbf{\stopmethod}} & \multicolumn{4}{c}{\textbf{AgentStop}} & \multicolumn{4}{c}{\textbf{Duration}} \\
\cmidrule(lr){3-6} \cmidrule(lr){7-10} \cmidrule(lr){11-14}
& & \textbf{Recall} & \textbf{Prec.} & \textbf{Fired} & \textbf{Saved} & \textbf{Recall} & \textbf{Prec.} & \textbf{Fired} & \textbf{Saved} & \textbf{Recall} & \textbf{Prec.} & \textbf{Fired} & \textbf{Saved} \\
\midrule
Qwen3.5-9B & $5$ & $34.7$ & $88.2$ & $\mathbf{20.4}$ & $\mathbf{15.7}$ & $\mathbf{35.1}$ & $\mathbf{89.2}$ & $\mathbf{20.4}$ & $10.2$ & $34.7$ & $88.2$ & $\mathbf{20.4}$ & $10.9$ \\
 & $10$ & $\mathbf{51.0}$ & $\mathbf{84.6}$ & $\mathbf{31.2}$ & $\mathbf{29.6}$ & $41.7$ & $83.1$ & $26.0$ & $20.9$ & $41.7$ & $81.8$ & $26.4$ & $21.2$ \\
 & $15$ & $\mathbf{61.0}$ & $\mathbf{81.4}$ & $\mathbf{38.8}$ & $\mathbf{38.4}$ & $50.2$ & $78.3$ & $33.2$ & $28.6$ & $48.3$ & $78.1$ & $32.0$ & $26.5$ \\
 & $20$ & $\mathbf{66.4}$ & $\mathbf{78.5}$ & $\mathbf{43.8}$ & $\mathbf{44.0}$ & $54.1$ & $74.5$ & $37.6$ & $35.9$ & $51.7$ & $74.4$ & $36.0$ & $32.3$ \\
 & $25$ & $\mathbf{72.2}$ & $\mathbf{75.7}$ & $\mathbf{49.4}$ & $\mathbf{50.1}$ & $62.9$ & $73.1$ & $44.6$ & $43.8$ & $57.1$ & $71.2$ & $41.6$ & $38.3$ \\
\midrule
Qwen3.6-27B & $5$ & $\mathbf{30.5}$ & $\mathbf{76.1}$ & $\mathbf{13.4}$ & $\mathbf{20.4}$ & $21.0$ & $68.6$ & $10.2$ & $12.5$ & $19.8$ & $67.3$ & $9.8$ & $11.4$ \\
 & $10$ & $\mathbf{45.5}$ & $\mathbf{69.7}$ & $\mathbf{21.8}$ & $\mathbf{28.0}$ & $34.1$ & $64.0$ & $17.8$ & $24.3$ & $29.9$ & $60.2$ & $16.6$ & $22.1$ \\
 & $15$ & $\mathbf{58.1}$ & $\mathbf{66.4}$ & $\mathbf{29.2}$ & $\mathbf{37.5}$ & $43.7$ & $60.3$ & $24.2$ & $28.4$ & $40.7$ & $60.2$ & $22.6$ & $26.1$ \\
 & $20$ & $\mathbf{62.9}$ & $\mathbf{61.8}$ & $\mathbf{34.0}$ & $\mathbf{44.9}$ & $48.5$ & $55.1$ & $29.4$ & $31.7$ & $46.1$ & $56.6$ & $27.2$ & $29.6$ \\
 & $25$ & $\mathbf{68.3}$ & $\mathbf{58.8}$ & $\mathbf{38.8}$ & $\mathbf{49.0}$ & $54.5$ & $52.3$ & $34.8$ & $37.2$ & $50.9$ & $51.8$ & $32.8$ & $33.7$ \\
\midrule
Gemma4-31B & $5$ & $\mathbf{27.2}$ & $\mathbf{82.1}$ & $\mathbf{13.4}$ & $\mathbf{14.6}$ & $23.8$ & $78.7$ & $12.2$ & $10.3$ & $20.8$ & $75.0$ & $11.2$ & $10.0$ \\
 & $10$ & $38.1$ & $72.6$ & $21.2$ & $\mathbf{23.0}$ & $\mathbf{38.6}$ & $\mathbf{72.9}$ & $\mathbf{21.4}$ & $18.5$ & $32.7$ & $70.2$ & $18.8$ & $16.6$ \\
 & $15$ & $\mathbf{46.5}$ & $\mathbf{68.6}$ & $\mathbf{27.4}$ & $\mathbf{26.9}$ & $43.6$ & $66.7$ & $26.4$ & $22.0$ & $37.1$ & $64.7$ & $23.2$ & $20.5$ \\
 & $20$ & $\mathbf{54.0}$ & $\mathbf{64.9}$ & $\mathbf{33.6}$ & $\mathbf{30.9}$ & $48.0$ & $63.8$ & $30.4$ & $25.0$ & $48.0$ & $63.8$ & $30.4$ & $25.0$ \\
 & $25$ & $\mathbf{57.4}$ & $61.1$ & $\mathbf{38.0}$ & $\mathbf{32.3}$ & $51.5$ & $\mathbf{61.9}$ & $33.6$ & $27.5$ & $51.5$ & $\mathbf{61.9}$ & $33.6$ & $27.5$ \\
\bottomrule
\end{tabular}
\caption{Early-stop monitor performance across false-positive rate (FPR) budgets. \emph{Recall}: percentage of true failing runs flagged; \emph{Prec.}: percentage of true failures among aborted runs; \emph{Fired} / \emph{Saved}: percentage of aborted trajectories and reclaimed tokens. \emph{Duration} is a supervisor-free control given only the step count. Bold marks the best value of each metric per budget.}
\label{tab:earlystop}
\end{table*}

%% file: tables/tab_gemini.tex
\begin{table}[!t]
\centering
\small
\setlength{\tabcolsep}{2.5pt}
\begin{tabular}{@{} l cccc cccc @{}}
\toprule
\textbf{FPR} & \multicolumn{4}{c}{\textbf{FailFast}} & \multicolumn{4}{c}{\textbf{Duration}} \\
\cmidrule(lr){2-5} \cmidrule(lr){6-9}
& \textbf{Recall} & \textbf{Prec.} & \textbf{Fired} & \textbf{Saved} & \textbf{Recall} & \textbf{Prec.} & \textbf{Fired} & \textbf{Saved} \\
\midrule
$5$ & $\mathbf{22.7}$ & $\mathbf{69.8}$ & $\mathbf{10.6}$ & $\mathbf{16.0}$ & $17.8$ & $64.4$ & $9.0$ & $12.0$ \\
$10$ & $\mathbf{31.9}$ & $\mathbf{61.9}$ & $\mathbf{16.8}$ & $\mathbf{23.3}$ & $23.9$ & $56.5$ & $13.8$ & $17.7$ \\
$15$ & $\mathbf{36.8}$ & $\mathbf{54.5}$ & $\mathbf{22.0}$ & $\mathbf{29.5}$ & $26.4$ & $46.7$ & $18.4$ & $22.8$ \\
$20$ & $\mathbf{42.9}$ & $\mathbf{51.9}$ & $\mathbf{27.0}$ & $\mathbf{33.8}$ & $28.2$ & $41.1$ & $22.4$ & $26.7$ \\
$25$ & $\mathbf{44.8}$ & $\mathbf{46.8}$ & $\mathbf{31.2}$ & $\mathbf{37.1}$ & $32.5$ & $40.8$ & $26.0$ & $30.2$ \\
\bottomrule
\end{tabular}
\caption{Early-stop monitor performance on closed-weight Gemini 3 Flash. AgentStop is inapplicable as closed APIs do not expose internal token logprobs.}
\label{tab:gemini}
\end{table}

%% file: tables/tab_restart.tex
\begin{table}[!t]
\centering
\small
\setlength{\tabcolsep}{2.0pt}
\begin{tabular}{@{}llcccc@{}}
\toprule
\textbf{FPR} & \textbf{Method} & \textbf{FP} & \textbf{TP} & \textbf{Resolve} & \textbf{Token} \\
 & & \textbf{lost} & \textbf{rec.} & \textbf{rate} & \textbf{overhead} \\
\midrule
\multicolumn{6}{@{}l}{\textbf{Policy Model: Qwen3.5-9B} (vanilla resolve rate: $48.2$)} \\
\midrule
$-$    & \emph{SWE-PRM}              & $-$ & $-$ & $52.4\,(+4.2)$ & \$0.15/t. \\
\cmidrule{1-6}
$10$ & \emph{\stopmethod{}+\restartmethod} & $45.8$ & $21.7$ & $52.2\,(+4.0)$ & $+36.9$ \\
       & \emph{\stopmethod{}+Cold Restart}         & $45.8$ & $16.8$ & $50.8\,(+2.6)$ & $+12.1$ \\
\cmidrule{1-6}
$25$ & \emph{\stopmethod{}+\restartmethod} & $25.0$ & $19.0$ & $52.0\,(+3.8)$ & $+39.2$ \\
       & \emph{\stopmethod{}+Cold Restart}         & $33.3$ & $15.6$ & $49.8\,(+1.6)$ & $+12.1$ \\
\midrule
\multicolumn{6}{@{}l}{\textbf{Policy Model: Qwen3.6-27B} (vanilla resolve rate: $66.6$)} \\
\midrule
$-$    & \emph{SWE-PRM}              & $-$ & $-$ & $63.4\,(-3.2)$ & \$0.12/t. \\
\cmidrule{1-6}
$10$ & \emph{\stopmethod{}+\restartmethod} & $15.2$ & $27.6$ & $69.8\,(+3.2)$ & $+30.3$ \\
       & \emph{\stopmethod{}+Cold Restart}         & $36.4$ & $19.7$ & $67.2\,(+0.6)$ & $+17.8$ \\
\cmidrule{1-6}
$25$ & \emph{\stopmethod{}+\restartmethod} & $8.8$ & $28.9$ & $71.8\,(+5.2)$ & $+43.8$ \\
       & \emph{\stopmethod{}+Cold Restart}         & $27.5$ & $20.2$ & $66.8\,(+0.2)$ & $+18.2$ \\
\midrule
\multicolumn{6}{@{}l}{\textbf{Policy Model: Gemma4-31B} (vanilla resolve rate: $62.2$)} \\
\midrule
$-$    & \emph{SWE-PRM}              & $-$ & $-$ & $63.0\,(+0.8)$ & \$0.12/t. \\
\cmidrule{1-6}
$10$ & \emph{\stopmethod{}+\restartmethod} & $12.9$ & $24.4$ & $65.2\,(+3.0)$ & $+20.5$ \\
       & \emph{\stopmethod{}+Cold Restart}         & $16.1$ & $17.9$ & $64.0\,(+1.8)$ & $+7.2$ \\
\cmidrule{1-6}
$25$ & \emph{\stopmethod{}+\restartmethod} & $9.1$ & $17.5$ & $65.2\,(+3.0)$ & $+36.9$ \\
       & \emph{\stopmethod{}+Cold Restart}         & $19.5$ & $17.5$ & $63.6\,(+1.4)$ & $+9.8$ \\
\bottomrule
\end{tabular}
\caption{Performance of restart strategies on interrupted runs. \emph{FPR} = target false-positive rate threshold; \emph{FP lost} = failed runs that would have passed without being aborted; \emph{TP rec.} = recovered failing runs; \emph{Resolve rate} = final success rate (parentheses show gains over uncontrolled baseline); \emph{Total token overhead} denotes the net additional token spending after early-stop savings, as a percentage of standard single-sample spending. Ununitized values are in \%; \emph{/t.} = per task.}
\label{tab:restart}
\end{table}

%% file: tables/tab_arch.tex
\begin{table}[!t]
\centering
\small
\setlength{\tabcolsep}{6pt}
\begin{tabular}{@{} l cc cc @{}}
\toprule
\textbf{FPR} & \multicolumn{2}{c}{\textbf{Policy: Qwen3.5-9B}} & \multicolumn{2}{c}{\textbf{Policy: Qwen3.6-27B}} \\
\cmidrule(lr){2-3} \cmidrule(l){4-5}
 & \textbf{0.6B Mon.} & \textbf{4B Mon.} & \textbf{0.6B Mon.} & \textbf{4B Mon.} \\
\midrule
$5$  & \textbf{39.8} & 37.8 & \textbf{30.5} & 25.7 \\
$10$ & \textbf{55.2} & 51.7 & \textbf{45.5} & 41.3 \\
$15$ & \textbf{62.2} & 57.9 & \textbf{58.1} & 47.3 \\
$20$ & \textbf{65.6} & 65.3 & \textbf{62.9} & 54.5 \\
$25$ & 69.1 & \textbf{69.5} & \textbf{68.3} & 58.7 \\
\bottomrule
\end{tabular}
\caption{Influence of monitor backbone size (\texttt{Qwen3-0.6B} vs.\ \texttt{4B}) on early-stop recall across different policy models (\texttt{Qwen3.5-9B} and \texttt{Qwen3.6-27B}) (all values in \%).}
\label{tab:arch}
\end{table}

%% file: tables/tab_input_ablation.tex
\begin{table}[!t]
\centering
\small
\setlength{\tabcolsep}{1.6pt}
\begin{tabular}{@{} l cccc cccc @{}}
\toprule
\textbf{FPR} & \multicolumn{4}{c}{\textbf{Policy: Qwen3.5-9B}} & \multicolumn{4}{c}{\textbf{Policy: Qwen3.6-27B}} \\
\cmidrule(lr){2-5} \cmidrule(l){6-9}
 & \textbf{Ours (I+W)} & \textbf{+S} & \textbf{$-$W} & \textbf{$-$I} & \textbf{Ours (I+W)} & \textbf{+S} & \textbf{$-$W} & \textbf{$-$I} \\
\midrule
$5$ & $\mathbf{39.8}$ & $34.4$ & $39.0$ & $35.5$ & $30.5$ & $\mathbf{31.1}$ & $22.8$ & $26.3$ \\
$10$ & $\mathbf{55.2}$ & $49.0$ & $46.7$ & $51.0$ & $\mathbf{45.5}$ & $44.9$ & $35.9$ & $38.3$ \\
$15$ & $\mathbf{62.2}$ & $59.1$ & $51.4$ & $58.3$ & $\mathbf{58.1}$ & $55.1$ & $44.9$ & $48.5$ \\
$20$ & $65.6$ & $\mathbf{67.6}$ & $56.8$ & $62.2$ & $\mathbf{62.9}$ & $60.5$ & $50.3$ & $57.5$ \\
$25$ & $69.1$ & $\mathbf{69.9}$ & $62.9$ & $67.2$ & $\mathbf{68.3}$ & $64.1$ & $56.3$ & $65.3$ \\
\bottomrule
\end{tabular}
\caption{Ablation of monitor input components on early-stop recall across FPR budgets (all values in \%). I = \texttt{[ISSUE]} (problem statement), W = \texttt{[WINDOW]} (recent detailed execution steps), S = \texttt{[STATE]} (abstract overview from step 1).}
\label{tab:input_ablation}
\end{table}

%% file: tables/tab_restart_patience.tex
\begin{table}[!t]
\centering
\small
\setlength{\tabcolsep}{2.5pt}
\begin{tabular}{@{} l ccc ccc @{}}
\toprule
\textbf{Strategy} & \multicolumn{3}{c}{\textbf{10\% FPR Budget}} & \multicolumn{3}{c}{\textbf{25\% FPR Budget}} \\
\cmidrule(lr){2-4} \cmidrule(l){5-7}
 & \textbf{FP} & \textbf{TP} & \textbf{Resolve} & \textbf{FP} & \textbf{TP} & \textbf{Resolve} \\
 & \textbf{lost} & \textbf{rec.} & \textbf{rate} & \textbf{lost} & \textbf{rec.} & \textbf{rate} \\
\midrule
Immediate & $21.2$ & $28.9$ & $69.6\,(+3.0)$ & $18.8$ & $22.8$ & $68.8\,(+2.2)$ \\
Wait (Ours) & $15.2$ & $27.6$ & $\mathbf{69.8}\,(+3.2)$ & $8.8$ & $28.9$ & $\mathbf{71.8}\,(+5.2)$ \\
\bottomrule
\end{tabular}
\caption{Influence of edit-completion patience (delaying termination until post-abort code edits settle to avoid incomplete overlays) on restart performance on \texttt{Qwen3.6-27B} (all values in \%). \emph{FP lost} = failed runs that would have passed without being aborted; \emph{TP recovery} = recovered failing runs; \emph{Immediate} halts right at the abort signal; \emph{Wait} (Ours) waits for the first post-abort edit and cuts after five edit-free steps.}
\label{tab:restart_patience}
\end{table}

%% file: sections/conclusion.tex
\section{Conclusion and Future Work}
\label{sec:conclusion}
We present \stopmethod--\restartmethod, a two-stage controller for early trajectory termination and edit-preserving recovery in autonomous software issue resolution.
Our framework early-terminates likely-to-fail trajectories via a lightweight monitor (\stopmethod), saving up to $20.4\%$ of execution tokens at a $5\%$ FPR budget (i.e., $\times 0.95$ base performance), and extracts partial code-edit overlays to guide same-model retries (\restartmethod).
By distilling constructive artifacts from interrupted runs instead of using expensive oversampling, our approach boosts resolution rates (up to $+5.2\%$) at a modest $20.5\%$--$43.8\%$ token overhead.
Our evaluation is limited to SWE-bench Verified with mini-swe-agent, so the findings may not generalize to other software engineering tasks or agent frameworks.
Generalizability across policy models is supported by consistent results across multiple model scales and families, including a closed-API model.
Future work includes: (1)~training higher-accuracy monitors for greater token savings; and (2)~extracting broader non-anchoring artifacts to preserve success on misclassified runs while turning more failing trajectories into successes.

%% file: sections/appendix.tex
\twocolumn[{\centering{\LARGE\bfseries Technical Supplement}\par\vspace{1.2em}}]
\suppressfloats[t]
\section*{Overview}

This supplement provides material supporting the main paper.
Appendix~\ref{sec:supp:significance} reports statistical significance tests and confidence
intervals for the main results. Appendix~\ref{sec:supp:ablations} presents the three
\stopmethod{} ablations referenced in the Experiments section (observation-window size,
progress-target heads, and training seed count), together with the \stopmethod{} monitors
whose abort decisions drive the \restartmethod{} experiments.
Appendix~\ref{sec:supp:prompt} gives the verbatim prompt through which
\restartmethod{} presents the inherited edits to the agent. Appendix~\ref{sec:supp:repro}
documents compute infrastructure and hyperparameter ranges, and
Appendix~\ref{sec:appendix:implementation} collects extended implementation details.
Numbering continues the main paper's: Tables~1--6 are found there, while
Tables~7--12 appear in this supplement.

\section{Statistical Significance of the Main Results}
\label{sec:supp:significance}

All comparisons in the paper are paired at the trajectory level on the same 500 evaluation
runs, so we test significance with exact two-sided McNemar tests on the discordant pairs;
confidence intervals are nonparametric bootstrap percentiles over trajectories
(10{,}000 resamples).

Table~\ref{tab:significance_earlystop} applies the paired test to the early-stop comparison
of Table~\ref{tab:earlystop} at all five budgets: \stopmethod{} aborts significantly more
truly-failing runs than AgentStop and Duration in 25 of 30 policy$\times$budget comparisons
and is never significantly worse; the insignificant cells correspond to the near-ties
visible in Table~\ref{tab:earlystop}. Table~\ref{tab:significance_restart} shows that the
\stopmethod{}+\restartmethod{} resolve gains of Table~\ref{tab:restart} are statistically
significant for \emph{every} policy and budget ($p\leq9.4\times10^{-3}$; strongest
$4.2\times10^{-5}$), with bootstrap 95\% CIs that exclude zero throughout; against the
cold-restart control the advantage is significant on Qwen3.6-27B and, pooled over the three
policies, at both budgets.

The \emph{Fix}/\emph{Brk} counts also make the accounting of Table~\ref{tab:restart}
explicit: there, \emph{FP lost} is the share of aborted would-pass runs whose restart fails
to resolve, and \emph{TP rec.}\ the share of aborted would-fail runs whose restart resolves.
\emph{Fix} and \emph{Brk} are exactly their numerators (e.g.\ for the 9B at the 10\%
budget, $11$ of $24$ aborted would-pass runs are lost $=45.8\%$). This is also why
\emph{FP lost} falls as the budget loosens: the aborted would-pass set grows (24 to 60 for
the 9B), while a larger share of it survives the restart.

By the same test, SWE-PRM yields no improvement significant at $\alpha{=}0.05$ on any
policy: 9B $63/42$ discordant, $p{=}0.050$ (marginal); 27B $24/40$ (a net decrease),
$p{=}0.060$; Gemma4 $40/36$, $p{=}0.73$. Each policy is paired against the baseline
behind its Table~\ref{tab:restart} cell, and its larger discordant sets (it intervenes in
every run) dilute comparable net gains.

\input{tables/tab_significance_earlystop.tex}
\input{tables/tab_significance_restart.tex}

\section{\stopmethod{} Ablations and the \stopmethod{} Monitors Behind the Restart Experiments}
\label{sec:supp:ablations}

\paragraph{Influence of Observation Window Size.}

\input{tables/tab_kwindow.tex}

To determine the necessary historical context, we ablate the observation-window size $k$
(recent steps in \texttt{[WINDOW]}) in Table~\ref{tab:kwindow}.
A window that is too narrow ($k=2$) lacks sufficient execution context and degrades
recall, while $k=8$ and $k=16$ are within evaluation noise of each other at most budgets.
We adopt $k=8$: doubling the window doubles the monitor's input cost without a dependable
accuracy gain.

\paragraph{Influence of Progress Targets (F2P/P2P).}
\label{para:p2p}

\input{tables/tab_p2p_ablation.tex}

To evaluate auxiliary progress supervision, we retrain the monitor backbone with one head
(\emph{V Only}), with two heads (\emph{V+F2P}), and with three heads (\emph{V+F2P+P2P}).
To isolate the training objective, every arm is scored on the value head alone at the
same fixed checkpoint (Table~\ref{tab:p2p_ablation}). The deployed monitors of
Tables~\ref{tab:earlystop} and~\ref{tab:restart_monitors} are different training runs
scored with the full calibrated combination, so absolute recalls should be compared
within this table only.
The results demonstrate that multi-task auxiliary training with the dense F2P target
provides positive representation alignment, enriching the $V$ head and significantly
boosting its failure detection recall (e.g., $+6.2\%$ at $5\%$ FPR).
In contrast, further incorporating P2P supervision offers no additional benefit, validating
\emph{V+F2P} as the optimal training objective.

\paragraph{Influence of Training Seed Count.}

\input{tables/tab_seed_count.tex}

To examine dataset scaling, we ablate the number of policy trajectories per issue $N$ used
to train the monitor (Table~\ref{tab:seed_count}).
With modest sampling the monitor is clearly weaker and adding a few more runs barely
helps ($N{=}3$ and $N{=}6$ perform similarly), whereas expanding to $N{=}11$ yields consistent improvements across budgets (up to
${\sim}10$ points at mid budgets) by exposing the monitor to a far richer diversity of
failure modes.
Since trajectory collection is a one-time upfront cost, scaling to $N{=}11$ provides a
worthwhile investment for deployment performance.

\paragraph{The \stopmethod{} Monitors Behind the Restart Experiments.}
\label{sec:supp:restart_monitors}

\input{tables/tab_restart_monitors.tex}

Table~\ref{tab:restart_monitors} reports the \stopmethod{} monitors whose abort decisions
the restart experiments consume: each policy gets a monitor trained on that policy's own
trajectories, collected exactly as in the main setup (11 independent runs, seeds 0--10,
over the 500 instances). This makes the quality of the abort signal behind
Table~\ref{tab:restart} explicit. By contrast, Table~\ref{tab:earlystop} evaluates a
single monitor, trained on 27B trajectories and applied unchanged to every policy.

\section{How \restartmethod{} Presents the Overlay to the Agent}
\label{sec:supp:prompt}

Communicating the inherited edits is itself delicate: the restart must be able to
\emph{use} the prior work without being anchored to a trajectory the monitor judged doomed.
The overlay is therefore exposed as a removable tool, never spliced into the source, and
the prompt explicitly warns that the inherited edits may be wrong. The following note is
appended to the task instruction of every restarted run that carries an overlay:

\begin{quote}\small\itshape
Note --- a previous attempt's edits are available as an overlay:\\
A previous attempt at this task EDITED the source; those changes are NOT applied --- the
source is pristine --- but are available as an overlay you can apply. That attempt was
stopped early because it was judged unlikely to finish, so its edits may be a correct fix,
an incomplete fix, or a wrong direction --- do NOT assume they are right. Your job is to
VERIFY them against the actual behavior described in the issue, KEEP what is correct, and
FIX or REMOVE what is wrong, then submit a correct patch. You can inspect/apply the
inherited edits with the \texttt{overlay} command (do NOT use
\texttt{git checkout}/\texttt{git reset}, which would also discard your own work):\\[2pt]
\texttt{overlay diff} \ \ \ \# show the inherited edits\\
\texttt{overlay status} \ \# are they currently applied? (they start NOT applied)\\
\texttt{overlay on} \ \ \ \ \# apply them onto the source\\
\texttt{overlay off} \ \ \ \# remove them again (keeps any new edits you made)\\[2pt]
Start by reproducing the issue, then apply the overlay (\texttt{overlay on}) to check
whether the inherited edits fix it; build from there. Producing the correct fix matters
more than keeping these edits.
\end{quote}

The overlay starts \emph{off} and the note directs the agent to reproduce the issue
before applying it, so verification is anchored to the task's pristine state rather than
to the inherited edits. Two further safeguards keep self-checking trustworthy: the
restart executes in the instance's prepared test-bed environment rather than a bare
interpreter, and we opt \emph{not} to inject auto-generated reproduction scripts, to
avoid the risk that a flawed reproduction becomes a false oracle the agent over-trusts.

\section{Compute Infrastructure and Hyperparameter Ranges}
\label{sec:supp:repro}

\paragraph{Compute infrastructure.}
Policy trajectories were generated with SGLang~(v0.5.x)-served models on NVIDIA
H100~80GB, A100~40GB, RTX~A5000~24GB, and RTX~6000~Ada~48GB GPUs (nodes with
128--512\,GB RAM; Ubuntu~22.04; Docker or Apptainer containers); grading used the
official SWE-bench Docker harness. Monitors (0.6B backbone, LoRA) train on 2~GPUs in under 6~h with PyTorch~2.6,
transformers~5.9, peft~0.19; calibration and scoring use scikit-learn~1.7 / numpy~2.2; the
agent scaffold is mini-swe-agent~2.2.8. Monitor inference over the evaluation set takes
minutes.

\paragraph{Hyperparameter ranges.}
Monitor training: frozen 0.6B backbone with LoRA $r{=}16$, peak
LR $10^{-4}$, cosine schedule over 1500 optimizer steps, checkpoints every 250 steps
($\{500,\dots,1500\}$),
weight decay $0.01$, LoRA dropout $0.05$, auxiliary progress-head weight $0.5$. Abort-rule
selection searches step-fraction floors $\{0.20,\dots,0.50\}$ (fractions of the fixed
$T{=}100$ budget, so a floor of $0.20$ is the main text's $\phi_{\min}$ of $20$ steps), score thresholds
$0.30$--$0.96$ (step $0.02$), cumulative vote counts $M\in[1,12]$ / consecutive-run lengths
$K\in[1,8]$, and calibrator regularization $C\in\{0.1,0.3,1,3\}$. Ablated design choices:
observation window $k\in\{2,4,8,16\}$; training runs per instance $N\in\{3,6,11\}$; monitor
backbone $\{0.6\mathrm{B},4\mathrm{B}\}$; input sections (Table~\ref{tab:input_ablation}).

\paragraph{Runs and variation.}
Every reported rate pools 100 held-out instances $\times$ 5 runs (500 trajectories); runs
are generated under fixed, distinct sampling seeds passed to the serving engine
(deterministic inference enabled where supported), and the same seeded trajectories are
reused across all methods, so comparisons are paired. The 350/50/100 instance split is
\emph{instance-grouped}, seeded, and frozen to JSON: all 11 runs of an instance stay on
the same side, so no instance leaks across train/validation/test.

\paragraph{Baseline implementations.}
AgentStop is our reimplementation of the authors' released code
(\url{github.com/brave-experiments/AgentStop}). SWE-PRM has no code release and is
reproduced from its paper, \emph{When Agents go Astray: Course-Correcting SWE Agents
with PRMs}. The \texttt{[STATE]} analysis is likewise reimplemented from scratch from
\emph{Process-Centric Analysis of Agentic Software Systems}.

\section{Extended Implementation Details}
\label{sec:appendix:implementation}

\paragraph{Monitor input serialization.}
Beyond the \texttt{[ISSUE]}+\texttt{[WINDOW]} serialization of the main paper, three
details matter for the input ablation (Table~\ref{tab:input_ablation}).
\emph{(i)} Each prefix is capped at $4{,}096$ tokens, enforced as a character budget at
$3.5$ characters per token; we trim observation head/tail first, then shrink the window,
preserving each step's thought and command and the pinned patch step. \emph{(ii)} Within \texttt{[WINDOW]}, each step shows the
agent's \emph{thought}, its \emph{command} (code edits receive a larger character budget;
read-only commands are clipped to one line), a classified \emph{error} tag, a
\emph{self-test} flag, a \emph{patch?} flag, and the head/tail-truncated
\emph{observation}. \emph{(iii)} \texttt{[STATE]} is a compact structural summary of the run so far (phase
sequence, loop and backtrack counts, progress $t/T$), computed with our from-scratch
reimplementation of the trajectory analysis of \emph{Process-Centric Analysis of Agentic
Software Systems}; it is evaluated in the ablation and omitted from the deployed
monitor. The gold F2P/P2P coverage and the
final verdict are label fields and never enter the text.

\paragraph{Recipe and policy selection.}
Two training choices were searched on the validation split by recall at the budgeted
FPRs: the F2P weight $\lambda_f\in\{0,0.25,0.5\}$ ($0.5$ won) and the checkpoint (saved
every $250$ steps; chosen per arm by best recall at the $10\%$ budget: step~$500$ for
9B, step~$1500$ for 27B).

%% file: tables/tab_significance_earlystop.tex
\begin{table}[!t]
\centering
\small
\setlength{\tabcolsep}{3.0pt}
\begin{tabular}{@{} l c cc @{}}
\toprule
\textbf{Policy} & \textbf{FPR} & \textbf{vs.\ AgentStop} & \textbf{vs.\ Duration} \\
\midrule
\multirow{5}{*}{\shortstack[l]{Qwen3.5\\-9B}} & $5$ & $6/7$ \;($1.0$) & $7/7$ \;($1.0$) \\
 & $10$ & $27/3$ \;($\mathbf{8.4\times10^{-6}}$) & $28/4$ \;($\mathbf{1.9\times10^{-5}}$) \\
 & $15$ & $35/7$ \;($\mathbf{1.5\times10^{-5}}$) & $39/6$ \;($\mathbf{5.4\times10^{-7}}$) \\
 & $20$ & $37/5$ \;($\mathbf{4.4\times10^{-7}}$) & $41/3$ \;($\mathbf{1.6\times10^{-9}}$) \\
 & $25$ & $33/9$ \;($\mathbf{2.7\times10^{-4}}$) & $43/4$ \;($\mathbf{2.8\times10^{-9}}$) \\
\cmidrule(lr){1-4}
\multirow{5}{*}{\shortstack[l]{Qwen3.6\\-27B}} & $5$ & $17/1$ \;($\mathbf{1.4\times10^{-4}}$) & $19/1$ \;($\mathbf{4.0\times10^{-5}}$) \\
 & $10$ & $29/10$ ($\mathbf{3.4\times10^{-3}}$) & $34/8$ \;($\mathbf{6.9\times10^{-5}}$) \\
 & $15$ & $32/8$ \;($\mathbf{1.8\times10^{-4}}$) & $37/8$ \;($\mathbf{1.5\times10^{-5}}$) \\
 & $20$ & $30/6$ \;($\mathbf{7.0\times10^{-5}}$) & $31/3$ \;($\mathbf{7.7\times10^{-7}}$) \\
 & $25$ & $28/5$ \;($\mathbf{6.6\times10^{-5}}$) & $33/4$ \;($\mathbf{1.1\times10^{-6}}$) \\
\cmidrule(lr){1-4}
\multirow{5}{*}{\shortstack[l]{Gemma4\\-31B}} & $5$ & $17/10$ ($0.25$) & $16/3$ \;($\mathbf{4.4\times10^{-3}}$) \\
 & $10$ & $16/17$ ($1.0$) & $18/7$ \;($\mathbf{0.043}$) \\
 & $15$ & $17/11$ ($0.34$) & $23/4$ \;($\mathbf{3.1\times10^{-4}}$) \\
 & $20$ & $17/5$ \;($\mathbf{0.017}$) & $17/5$ \;($\mathbf{0.017}$) \\
 & $25$ & $15/3$ \;($\mathbf{7.5\times10^{-3}}$) & $15/3$ \;($\mathbf{7.5\times10^{-3}}$) \\
\bottomrule
\end{tabular}
\caption{Paired comparison of abort decisions on the truly-failing runs at matched false-positive budgets (\emph{FPR}, in \%): $a/b$ = the \emph{number} of failing runs aborted only by \stopmethod{} / only by the baseline, with the exact two-sided McNemar $p$ in parentheses (boldface: significant at $p<0.05$). \stopmethod{} is significantly better in 25 of 30 comparisons and never significantly worse; every insignificant cell corresponds to a near-tie visible in Table~\ref{tab:earlystop} (all methods at the tightest 9B budget, and AgentStop on Gemma4-31B at the 5--15\% budgets).}
\label{tab:significance_earlystop}
\end{table}

%% file: tables/tab_significance_restart.tex
\begin{table}[!t]
\centering
\small
\setlength{\tabcolsep}{1.0pt}
\begin{tabular}{@{} l c c c c c @{}}
\toprule
\textbf{Policy} & \textbf{FPR} & \textbf{$\Delta$Res.\ [95\% CI]} & \textbf{Fix/Brk} & \textbf{$p$ vs.\ vanilla} & \textbf{$p$ vs.\ cold} \\
\midrule
\multirow{2}{*}{\shortstack[l]{Qwen3.5\\-9B}} & $10$ & $+4.0\,[+1.6,+6.6]$ & $31/11$ & $\mathbf{2.9\times10^{-3}}$ & $0.31$ \\
 & $25$ & $+3.8\,[+1.2,+6.4]$ & $34/15$ & $\mathbf{9.4\times10^{-3}}$ & $0.15$ \\
\cmidrule(lr){1-6}
\multirow{2}{*}{\shortstack[l]{Qwen3.6\\-27B}} & $10$ & $+3.2\,[+1.2,+5.2]$ & $21/5$  & $\mathbf{2.5\times10^{-3}}$ & $\mathbf{0.029}$ \\
 & $25$ & $+5.2\,[+2.8,+7.6]$ & $33/7$  & $\mathbf{4.2\times10^{-5}}$ & $\mathbf{0.001}$ \\
\cmidrule(lr){1-6}
\multirow{2}{*}{\shortstack[l]{Gemma4\\-31B}} & $10$ & $+3.0\,[+1.2,+5.0]$ & $19/4$  & $\mathbf{2.6\times10^{-3}}$ & $0.33$ \\
 & $25$ & $+3.0\,[+0.8,+5.0]$ & $22/7$  & $\mathbf{8.1\times10^{-3}}$ & $0.32$ \\
\bottomrule
\end{tabular}
\caption{Statistical support for the \stopmethod{}+\restartmethod{} gains in Table~\ref{tab:restart} (\emph{FPR}, in \%). \emph{$\Delta$Res.}: resolve-rate change (in \%), with a nonparametric bootstrap 95\% CI over the 500 evaluation trajectories (10{,}000 resamples). \emph{Fix/Brk}: the \emph{number} of failing runs repaired vs.\ passing runs broken, i.e.\ the discordant pairs of an exact two-sided McNemar test against the vanilla (uncontrolled) runs of Table~\ref{tab:restart} ($p$ vs.\ vanilla; significant for every policy and budget; boldface: $p<0.05$). \emph{$p$ vs.\ cold}: the same test paired on the identical aborted runs against the cold-restart control; pooled over the three policies it reaches $p{=}8.8\times10^{-3}$ (10\% FPR) and $p{=}4.5\times10^{-4}$ (25\% FPR).}
\label{tab:significance_restart}
\end{table}

%% file: tables/tab_kwindow.tex
\begin{table}[!t]
\centering
\small
\setlength{\tabcolsep}{1.3pt}
\begin{tabular}{@{} l cccc cccc @{}}
\toprule
\textbf{FPR} & \multicolumn{4}{c}{\textbf{Policy: Qwen3.5-9B}} & \multicolumn{4}{c}{\textbf{Policy: Qwen3.6-27B}} \\
\cmidrule(lr){2-5} \cmidrule(l){6-9}
 & \textbf{$k{=}2$} & \textbf{$k{=}4$} & \textbf{Ours ($k{=}8$)} & \textbf{$k{=}16$} & \textbf{$k{=}2$} & \textbf{$k{=}4$} & \textbf{Ours ($k{=}8$)} & \textbf{$k{=}16$} \\
\midrule
$5$ & $35.1$ & $32.0$ & $39.8$ & $\mathbf{44.4}$ & $32.9$ & $28.7$ & $30.5$ & $\mathbf{37.1}$ \\
$10$ & $51.0$ & $49.0$ & $\mathbf{55.2}$ & $52.5$ & $41.9$ & $44.3$ & $45.5$ & $\mathbf{48.5}$ \\
$15$ & $59.8$ & $57.1$ & $\mathbf{62.2}$ & $61.4$ & $48.5$ & $55.1$ & $\mathbf{58.1}$ & $52.7$ \\
$20$ & $64.9$ & $59.8$ & $65.6$ & $\mathbf{66.0}$ & $53.9$ & $59.9$ & $\mathbf{62.9}$ & $59.3$ \\
$25$ & $68.0$ & $64.9$ & $69.1$ & $\mathbf{71.8}$ & $59.9$ & $65.3$ & $\mathbf{68.3}$ & $65.9$ \\
\bottomrule
\end{tabular}
\caption{Influence of the observation-window size $k$ (last-$k$ agent steps) on early-stop recall across FPR budgets (all values in \%). Ours uses $k{=}8$; the $k{=}8$ column reproduces the deployed configuration of Table~\ref{tab:earlystop}.}
\label{tab:kwindow}
\end{table}

%% file: tables/tab_p2p_ablation.tex
\begin{table}[!t]
\centering
\small
\setlength{\tabcolsep}{8pt}
\begin{tabular}{@{} c ccc @{}}
\toprule
\textbf{FPR} & \textbf{V Only} & \textbf{Ours (V+F2P)} & \textbf{V+F2P+P2P} \\
\midrule
$5$  & 35.5 & $\mathbf{41.7}$ & 37.8 \\
$10$ & 54.8 & $\mathbf{55.2}$ & 52.5 \\
$15$ & 62.2 & $\mathbf{63.7}$ & 63.3 \\
$20$ & 67.6 & $\mathbf{68.3}$ & 67.2 \\
$25$ & $\mathbf{72.2}$ & 70.3 & 71.4 \\
\bottomrule
\end{tabular}
\caption{Influence of progress target formulation on early-stop recall across FPR budgets on Qwen3.5-9B (all values in \%). \emph{V}: the monitor's value head; F2P / P2P: the Fail-to-Pass / Pass-to-Pass auxiliary progress heads.}
\label{tab:p2p_ablation}
\end{table}

%% file: tables/tab_seed_count.tex
\begin{table}[!t]
\centering
\small
\setlength{\tabcolsep}{2.9pt}
\begin{tabular}{@{} l ccc ccc @{}}
\toprule
\textbf{FPR} & \multicolumn{3}{c}{\textbf{Policy: Qwen3.5-9B}} & \multicolumn{3}{c}{\textbf{Policy: Qwen3.6-27B}} \\
\cmidrule(lr){2-4} \cmidrule(l){5-7}
 & \textbf{$N{=}3$} & \textbf{$N{=}6$} & \textbf{Ours ($N{=}11$)} & \textbf{$N{=}3$} & \textbf{$N{=}6$} & \textbf{Ours ($N{=}11$)} \\
\midrule
$5$ & $33.6$ & $34.0$ & $\mathbf{39.8}$ & $29.3$ & $29.3$ & $\mathbf{30.5}$ \\
$10$ & $47.9$ & $44.8$ & $\mathbf{55.2}$ & $40.7$ & $41.9$ & $\mathbf{45.5}$ \\
$15$ & $57.1$ & $56.0$ & $\mathbf{62.2}$ & $49.7$ & $48.5$ & $\mathbf{58.1}$ \\
$20$ & $64.1$ & $62.9$ & $\mathbf{65.6}$ & $55.7$ & $52.1$ & $\mathbf{62.9}$ \\
$25$ & $67.6$ & $68.3$ & $\mathbf{69.1}$ & $60.5$ & $57.5$ & $\mathbf{68.3}$ \\
\bottomrule
\end{tabular}
\caption{Influence of policy training seed count $N$ (trajectories per issue) on early-stop recall across FPR budgets (all values in \%). Ours uses $N{=}11$; the $N{=}11$ column reproduces the deployed configuration of Table~\ref{tab:earlystop}.}
\label{tab:seed_count}
\end{table}

%% file: tables/tab_restart_monitors.tex
\begin{table}[!t]
\centering
\small
\setlength{\tabcolsep}{4.5pt}
\begin{tabular}{@{} l l cccc @{}}
\toprule
\textbf{Policy Model} & \textbf{FPR} & \textbf{Recall} & \textbf{Prec.} & \textbf{Fired} & \textbf{Saved} \\
\midrule
Qwen3.5-9B & $5$ & $39.8$ & $89.6$ & $23.0$ & $27.4$ \\
 & $10$ & $55.2$ & $85.6$ & $33.4$ & $41.8$ \\
 & $15$ & $62.2$ & $81.7$ & $39.4$ & $47.3$ \\
 & $20$ & $65.6$ & $78.0$ & $43.6$ & $50.5$ \\
 & $25$ & $69.1$ & $74.9$ & $47.8$ & $57.5$ \\
\midrule
Qwen3.6-27B & $5$ & $30.5$ & $76.1$ & $13.4$ & $20.4$ \\
 & $10$ & $45.5$ & $69.7$ & $21.8$ & $28.0$ \\
 & $15$ & $58.1$ & $66.4$ & $29.2$ & $37.5$ \\
 & $20$ & $62.9$ & $61.8$ & $34.0$ & $44.9$ \\
 & $25$ & $68.3$ & $58.8$ & $38.8$ & $49.0$ \\
\midrule
Gemma4-31B & $5$ & $25.9$ & $76.6$ & $12.8$ & $22.8$ \\
 & $10$ & $41.3$ & $71.6$ & $21.8$ & $34.0$ \\
 & $15$ & $52.9$ & $69.0$ & $29.0$ & $43.0$ \\
 & $20$ & $60.3$ & $65.1$ & $35.0$ & $46.3$ \\
 & $25$ & $66.7$ & $62.1$ & $40.6$ & $54.2$ \\
\bottomrule
\end{tabular}
\caption{Performance of the \emph{policy-specific} early-stop monitors that produced the abort decisions consumed by \restartmethod{} (Table~\ref{tab:restart}). Unlike Table~\ref{tab:earlystop}, where a single monitor trained on 27B trajectories supervises every policy, each monitor here is trained on its own policy's trajectories, the strongest abort signal available for restart. The Qwen3.6-27B block is identical to Table~\ref{tab:earlystop} by construction. \emph{Recall} / \emph{Prec.}: truly-failing runs aborted / aborted runs that are true failures; \emph{Fired} / \emph{Saved}: trajectories aborted / tokens reclaimed (all values in \%). These policy-specific figures feed the restart experiments only; the headline token savings quoted in the abstract are those of the transferred monitor in Table~\ref{tab:earlystop}.}
\label{tab:restart_monitors}
\end{table}